\documentclass[aps,pra,reprint,superscriptaddress,longbibliography]{revtex4-1}

\usepackage{epstopdf}
\usepackage{amsmath,amssymb,graphicx,bm}
\usepackage{graphicx}
\usepackage{dcolumn}
\usepackage{bm}

\begin{document}


\title{Anomalous diffraction in hyperbolic materials}

\author{Alessandro Alberucci}
\email[]{alessandro.alberucci@tut.fi}
\affiliation{Optics Laboratory, Tampere University of Technology, FI-33101 Tampere, Finland}

\author{Chandroth P. Jisha} 
\affiliation{Centro de F\'{\i}sica do Porto, Faculdade de Ci\^encias, Universidade do Porto, 
4169-007 Porto, Portugal}

\author{Allan D. Boardman}
\email[]{A.D.boardman@salford.ac.uk}
\affiliation{Joule Material \& Physics Research Centre, University of Salford, Salford M5 4WT, UK}

\author{Gaetano Assanto}
\affiliation{Optics Laboratory, Tampere University of Technology, FI-33101 Tampere, Finland}
\affiliation{N\textsl{oo}EL - Nonlinear Optics and OptoElectronics Lab, University ``Roma Tre'', IT-00146 Rome, Italy}
\affiliation{CNR-ISC, Institute for Complex Systems, IT-00185 Rome, Italy}

\date{\today}

\begin{abstract}
We demonstrate that light is subject to anomalous (i.e., negative) diffraction when propagating in the presence of hyperbolic dispersion.
We show that light propagation in hyperbolic media resembles the dynamics of a quantum particle of negative mass moving in a two-dimensional potential. The negative effective mass implies time reversal if the medium is homogeneous. Such property paves the way to diffraction compensation, spatial analogue of dispersion compensating fibers in the temporal domain. 
At variance with materials exhibiting standard elliptic dispersion, in inhomogeneous hyperbolic materials light waves are pulled towards regions with a lower refractive index. In the presence of a Kerr-like optical response, bright (dark) solitons are supported by a negative (positive) nonlinearity.   
\end{abstract}


\maketitle
\section{Introduction}
The linear propagation of electromagnetic waves in homogeneous media can be fully described by plane wave eigen-solutions, resulting in a dispersion curve linking the wave vector $\bm{k}$ to the frequency (pulsation) $\omega$. Macroscopically, the dispersion (or existence) curve $\omega(\bm{k})$ markedly depends on the  constitutive equations \cite{Kong:1990}, which describe the material response to electromagnetic fields. In the simplest case of isotropic media (e.g., vacuum), the isofrequency surfaces $\omega=\omega(\bm{k})$ are spheres in the transformed $\bm{k}$-space. When the medium is anisotropic, the isofrequency surfaces are ellipsoids, i.e., they maintain the same topology as long as all the eigenvalues of both the permittivity tensor $\bm{\epsilon}$ and the magnetic permeability $\bm{\mu}$ are positive. The dispersion topology  drastically changes if the eigenvalues have opposite signs, leading to a hyperbolic (aka indefinite) dispersion with isofrequency 
hyperboloids \cite{Smith:2003,Poddubny:2013,Ferrari:2015}. \\
Hyperbolic materials (HMs) have attracted a great deal of attention in optics owing to their unique properties. Most notably, HMs can support propagating plane waves regardless of the transverse wave vector, thus allowing for sub-wavelength optical resolution via the so-called hyperlens effect \cite{Jacob:2006}. HMs can exhibit negative refraction, with much smaller sensitivity to dispersion and losses than negative refractive index materials (NRIM) \cite{Smith:2004,Yao:2008}. When inserted in planar waveguides they can even emulate the main features of NRIM \cite{Podolskiy:2005}; hyperbolic cores also strongly affects the existence region of guided and leaky modes \cite{Boardman:2015}. In addition, HMs exhibit a highly increased density of photonic states, with a consequent enhancement of the Purcell factor \cite{Noginov:2010,Jacob:2012,Krishnamoorthy:2012}. HMs also allow for the possibility of realizing exotic nanocavities \cite{Yang:2012}, improved slot waveguides \cite{He:2012}, nanolithography \cite{Ishii:2013}, novel phase-matching configurations in nonlinear optics \cite{Duncan:2015}, the engineering of photon-mediated heat exchange \cite{Biehs:2015} and ultra-sensitive biosensors \cite{Sreekanth:2016}. HMs have been suggested as an ideal workbench for exploring optical analogues of relativistic and cosmological phenomena \cite{Smolyaninov:2010,Smolyaninov:2015}. While the first HMs were metamaterials consisting of either a stack of subwavelength metal/dielectric elements or a web of metallic nanowires embedded in a dielectric \cite{Poddubny:2013}, hyperbolic dispersion  was later discovered in nature \cite{Sun:2014,Esslinger:2014,Narimanov:2015}, including, e.g., magnetized plasma \cite{Zhang:2011,Abelson:2015}. An exhaustive review about natural hyperbolic materials is provided in Ref.~\cite{Korzeb:2015}. \\
In non-magnetic materials ($\bm{\mu}=\mu_0 \bm{I}$), HMs can be studied as an extreme case of anisotropy \cite{Elser:2006,Krishnamoorthy:2012,Drachev:2013}, with metallic or dielectric responses depending on the propagation direction and/or the input polarization. Here, inspired by such consideration, we investigate the propagation of electromagnetic waves in hyperbolic media generalizing the treatment previously introduced with reference to standard (i.e., elliptically dispersive) media \cite{Alberucci:2011}. The approach we undertake allows us to outline a direct analogy between light propagation and the evolution of particles with a negative inertial mass; however, in contrast to previous studies, an effective negative mass stems from neither a periodic band structure \footnote{For hyperbolic metamaterials a periodic subwavelength structure is required, but HMs macroscopically behave and can be treated as homogeneous media  whenever homogeneization theory applies.} for excitations close to the Bragg condition \cite{Rosencher:2002,Luo:2002,Wimmer:2013} nor nonlinear effects associated with specific wave profiles in space \cite{DiMei:2016}. Beyond a simpler understanding  of well-known effects such as negative refraction and hyperlensing in hyperbolic media, we introduce a spatial analogue of dispersion-compensation in fibers, allowing for the perfect reconstruction of an electromagnetic signal regardless of its shape. Finally, in the nonlinear limit we predict that self-focusing (self-defocusing) occurs when the refractive index decreases (increases) with the field intensity. Thus, bright (dark) spatial solitons are supported by a negative (positive) nonlinearity. \\

\section{Maxwell's equations in uniaxial media}
\label{sec:maxwell}
\begin{figure}
\includegraphics[width=0.5\textwidth]{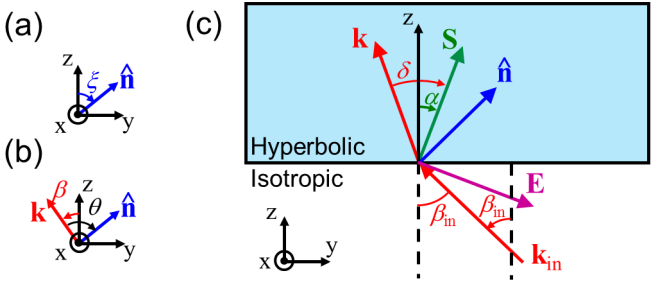}
\caption{\label{fig:sketch} (Color Online) Geometry of the wave-matter interaction detailed in the text. (a) The optic axis $\hat{n}$ lies in the plane $yz$ at an angle $\xi$  with axis $z$. (b) The wave vector $\bm{k}$ forms an angle $\beta$ with $z$, and an angle $\theta$ (positive in the picture) with $\hat{n}$. (c) Sketch of a beam impinging on an HM at an angle $\beta_\mathrm{in}<0$. In the hyperbolic material the Poynting vector is at an angle $\alpha$ with respect to $z$. In general, $\alpha\neq\beta$ due to walk-off $\delta$ between wave and Poynting vectors ($\delta>0$ in the example). All the angles are defined positive for clockwise rotations as seen by the viewer.}
\end{figure}

Let us start by considering the propagation of monochromatic electromagnetic waves (fields $\propto e^{-i\omega t}$) in non-magnetic dielectrics with a uniaxial response.  For an optic axis $\hat{n}$ lying in the plane $yz$, the relative permittivity tensor $\bm{\epsilon}$ reads
\begin{equation}  \label{eq:dielectric_tensor}
  \bm{\epsilon} = \left ( \begin{array} {ccc}
  \epsilon_{xx} & 0 & 0 \\
  0 & \epsilon_{yy} & \epsilon_{yz} \\
	 0 & \epsilon_{zy} & \epsilon_{zz} \\
 \end{array} \right)
\end{equation} 
where, in general, all the elements $\epsilon_{ij}$ are point-wise functions. In writing Eq.~\eqref{eq:dielectric_tensor} we assumed  the medium to be local, i.e., with $\bm{\epsilon}$ independent of the wave vector $\bm{k}$ \cite{Silveirinha:2009}. Naming $\epsilon_\bot$ and $\epsilon_\|$ the eigenvalues of $\bm{\epsilon}$ corresponding to electric fields normal and parallel to the optic axis $\hat{n}$, respectively, we get $\epsilon_{ij}=\epsilon_\bot \delta_{ij} + \epsilon_a n_i n_j \ (i,j=x,y,z)$, with $\epsilon_a=\epsilon_\|-\epsilon_\bot$ the optical anisotropy and $\xi$ the angle between $\hat{n}$ and $\hat{z}$, such that $\hat{n}= (0, \sin\xi , \cos\xi )$ [see Fig.~\ref{fig:sketch}(a)]. 
The Maxwell's equations in the absence of sources read
\begin{align} 
   &   \nabla\times \bm{E}= i\omega \mu_0 \bm{H}, \label{eq:maxwell_1} \\
   &   \nabla\times \bm{H}= -i\omega \epsilon_0 \bm{\epsilon}\cdot \bm{E} -i \omega \bm{P}_\mathrm{NL}, \label{eq:maxwell_2}
\end{align}
with $\bm{P}_\mathrm{NL}$ the nonlinear contribution to the electric polarization. For the sake of simplicity we refer to a (1+1)D geometry by considering an $x$-invariant system and setting $\partial_x=0$. In this limit, ordinary and extraordinary waves are decoupled even in the non-paraxial regime. Hereby we deal with extraordinary waves, i.e., with electric field oscillating in the plane $yz$. Equations~(\ref{eq:maxwell_1}-\ref{eq:maxwell_2}) projected on a Cartesian reference system yield \cite{Alberucci:2011}
\begin{align}
  &E_z=-\frac{i}{\omega \epsilon_0 \epsilon_{zz}} \frac{\partial H_x}{\partial y} - \frac{\epsilon_{zy}}{\epsilon_{zz}} E_y - \frac{1}{\epsilon_0 \epsilon_{zz}} P_\mathrm{NL,z}, \label{eq:Ez} \\
	&\frac{\partial H_x}{\partial z} + \frac{\epsilon_{yz}}{\epsilon_{zz}} \frac{\partial H_x}{\partial y} =  -i\omega \epsilon_0 \left( \epsilon_{yy} - \frac{\epsilon_{yz}\epsilon_{zy}}{\epsilon_{zz}} \right) E_y \nonumber  \\ & -i\omega \epsilon_0 \left( P_\mathrm{NL,y} - \frac{\epsilon_{yz}}{\epsilon_{zz}} P_\mathrm{NL,z}  \right), \label{eq:maxwell_y_2} \\
	&\frac{\partial E_y}{\partial z} + \frac{\epsilon_{zy}}{\epsilon_{zz}} \frac{\partial E_y}{\partial y} = -i\omega \mu_0 H_x - \frac{i}{\omega\epsilon_0 \epsilon_{zz}} \frac{\partial^2 H_x}{\partial y^2} - \frac{1}{\epsilon_0\epsilon_{zz}} \frac{\partial P_\mathrm{NL,z}}{\partial y}, \label{eq:maxwell_x_2} 
\end{align}
where we neglected the spatial derivatives of $\epsilon_{ij}$ with respect to $y$. Furthermore, we assumed that electric fields oscillating in the plane $yz$ do not to couple with nonlinear polarization components along $x$. The system of Eqs.~(\ref{eq:Ez}-\ref{eq:maxwell_x_2}) rules light propagation in both paraxial and non-paraxial regimes, regardless of the nature of $\bm{P}_\mathrm{NL}$, and is valid for both real (positive and/or negative) and  complex valued tensor elements $\epsilon_{ij}$. The first derivatives of the field with respect to $y$ on the LHS of Eqs.~(\ref{eq:maxwell_y_2}-\ref{eq:maxwell_x_2}) rule spatial walk-off, i.e., in general, the non-parallelism of the wave vector $\bm{k}$ to the Poynting vector $\bm{S}$.\\
We examine the case of a lossless medium, i.e., with Hermitian dielectric tensor such that $\epsilon_{ij}=\epsilon_{ji}^*$. Additionally, if  $\bm{\epsilon}$ is purely real, the Poynting vector $\bm{S}$ forms an angle $\delta=\arctan{\left(\frac{\epsilon_{yz}}{\epsilon_{zz}}\right)}$ with the average wave vector of the wavepacket. The average wave vector is chosen parallel to $z$. The extraordinary refractive index of a plane wave with $\bm{k}$ at an angle $\theta$ with $\hat{n}$ is $n_e(\theta)=\sqrt{\epsilon_{yy} - \frac{\epsilon_{yz}^2}{\epsilon_{zz}}}$. Hereafter we consider configurations with negligible variations along $y$ of the walk-off $\delta$, and an optic axis $\hat{n}$ at an angle $\xi$ with $z$, such that $\theta=\xi-\beta$ and $\bm{k}\cdot\hat{z}=\left|\bm{k}\right|\cos\beta$ [see Fig.~\ref{fig:sketch}(b)]. Computing $E_y$ from \eqref{eq:maxwell_y_2} and substituting into \eqref{eq:maxwell_x_2} we find 

\begin{align}
      &\frac{\partial^2 H_x}{\partial z'^2} + D_y \frac{\partial^2 H_x}{\partial y'^2} + k_0^2 n_e^2 H_x	=  \nonumber \\
			& -i\omega  \frac{\partial \left( P_\mathrm{NL,y} - \frac{\epsilon_{yz}}{\epsilon_{zz}} P_\mathrm{NL,z} \right)}{\partial z'} + \frac{i\omega n_e^2}{\epsilon_{zz}}  \frac{\partial P_\mathrm{NL,z}}{\partial y'},  \label{eq:single_eq_H} 
\end{align}
where we  introduced the moving reference system $x^\prime y^\prime z^\prime$ 
\begin{align}
 &x^\prime=x, \\
 &y^\prime=y-\frac{\epsilon_{yz}}{\epsilon_{zz}} z ,   \\
 &z^\prime=z    \label{eq:shifted_ref_system}
\end{align}
and assumed $\partial_{z'}n_e\approx 0$ (i.e., small longitudinal variations of the index on the wavelength scale). Although Eq.~\eqref{eq:single_eq_H} remains valid in the non-paraxial regime (off-axis propagation and wavelength-size beams), the quantities $D_y$, $n_e$ and $\delta$ have a simple physical meaning (diffraction coefficient, extraordinary refractive index and walk-off angle, respectively) only when $\bm{k}$ is directed along $z$, that is, when $\xi=\theta$, due to the anisotropic response on the plane $yz$. When the average $\bm{k}$ is not parallel to $z$, the three quantities retrieve their physical interpretation after a rotation of the original framework $xyz$ around $x$.

\section{Negative diffraction}
The diffraction coefficient in Eq.~\eqref{eq:single_eq_H}
\begin{equation} \label{eq:diffraction_coefficient}
D_y(\theta)=\frac{n_e^2}{\epsilon_{zz}}=\frac{\epsilon_\bot\epsilon_\|}{\epsilon_{zz}^2}=\frac{\epsilon_\bot \epsilon_\|}{\left(\epsilon_\bot+{\epsilon_a} \cos^2\theta\right)^2}
\end{equation}
depends on the angle $\theta$ and governs (in the paraxial regime) the propagation of finite beams according to the spatial spectrum of $\tilde{H_x}$ \cite{Paul:2009}, i.e.,  
\begin{align}
 H_x(y,z)  &= e^{ik_0 n_0 z} \int_{-\infty}^\infty{ \tilde{H_x}(k_y,z=0) e^{ik_y y}} \nonumber \\
 & {e^{i\left. \frac{\partial k_z}{\partial k_y}\right|_{k_y=0} k_y z} e^{i\left. \frac{\partial^2 k_z}{\partial k_y^2}\right|_{k_y=0} k_y^2 z} dk_y}, \label{eq:plane_waves_expansion}
\end{align}
with $\left. \frac{\partial^2 k_z}{\partial k_y^2}\right|_{k_y=0}=-\frac{D_y}{k_0 n_0}$. Since we are interested in hyperbolic (indefinite) media featuring
\begin{equation}
  \epsilon_\bot \epsilon_\| <0, \label{eq:negative_hyp}
\end{equation}
it is apparent from Eq.~\eqref{eq:diffraction_coefficient} that, whenever $n_e(\theta)$ is real, the diffraction coefficient is negative, i.e., diffraction is anomalous even if the refractive index remains positive. Noteworthy, in hyperbolic materials satisfying \eqref{eq:negative_hyp}, either purely real (propagating waves) or purely imaginary (evanescent waves) wave vectors are permitted according to the sign of $n_e^2(\theta)$. We further need to distinguish type I and type II HMs corresponding to $\epsilon_\|<0$ and $\epsilon_\|>0$, respectively. Propagating waves exist when
\begin{align}
  &\ \text{Type I: } &|\theta| < \arccos{\sqrt{\frac{\epsilon_\bot}{|\epsilon_\|| + \epsilon_\bot}}},  \label{eq:typeI}\\
 &\ \text{Type II: } &\arccos{\sqrt{\frac{|\epsilon_\bot|}{\epsilon_\| + |\epsilon_\bot|}}}  < |\theta| < \frac{\pi}{2}. 
\label{eq:typeII}
\end{align}

\begin{figure}
\includegraphics[width=0.5\textwidth]{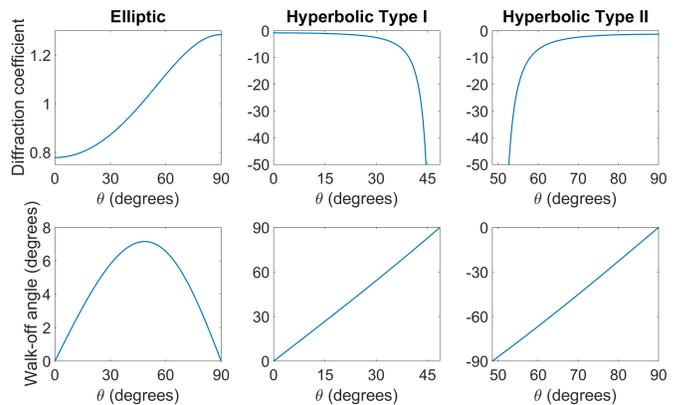}
\caption{\label{fig:comparison_diffraction} Diffraction coefficient $D_y$ (top) and walk-off angle $\delta$ (bottom) versus the angle $\theta$ between wave vector $\bm{k}$ and optic axis $\hat{n}$ in the presence of elliptic dispersion (left), hyperbolic dispersion of Type I (center) and Type II (right). Here $\left|n_\bot\right|=1.5$ and $\left|n_\|\right|=1.7$.}
\end{figure}

\begin{figure}
\includegraphics[width=0.5\textwidth]{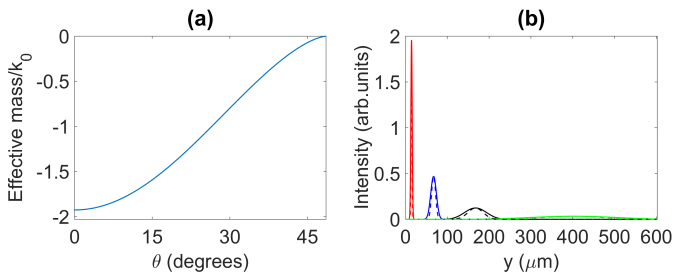}
\caption{\label{fig:effective_mass} Effective mass and transverse beam profile in a Type I hyperbolic material. (a) The effective mass scaled over the wave number (thus corresponding to $n_0/D_y$) versus angle $\theta$, for $\epsilon_\bot=2.25$ and $\epsilon_\|=-2.89$. (b) Diffracting beam profile computed in $z=20~\mu$m for $\theta=20^\circ$ (red line), $\theta=40^\circ$ (blue), $\theta=45^\circ$ (black) and $\theta=47^\circ$ (green), from left to right, respectively. Solid and dashed lines correspond  to the exact field and to the paraxial prediction \eqref{eq:plane_waves_expansion}, respectively. Here the input is a Gaussian of waist $2~\mu$m at wavelength 1064~nm, impinging normally on the material ($\beta_{in}=0$) with planar phase-front.}
\end{figure}

Figure~\ref{fig:comparison_diffraction} compares walk-off $\delta$ and diffraction coefficient $D_y$ versus $\theta$ for elliptic and hyperbolic dispersions, respectively. While diffraction  is  positive and finite in the elliptic case, $D_y$ is always negative in the hyperbolic case, consistently with Eq.~\eqref{eq:diffraction_coefficient}. Moreover, $D_y$  monotonically decreases (increases) in Type I (Type II) HMs, with a singularity when $n_e$ goes to infinity at the edge of the existence region for homogeneous plane waves. The walk-off angle monotonically increases when the dispersion is hyperbolic, remaining positive (negative) for Type I (II) materials and reaching an absolute maximum at $90^\circ$ when $n_e\rightarrow\infty$. 
The latter limit 
corresponds to the propagation of volume plasmon polaritons, as investigated in Ref.~\cite{Ishii:2013}.\\
Reverting back to the laboratory framework $xyz$, the single scalar equation \eqref{eq:single_eq_H} in the paraxial regime becomes
\begin{align}
      i\left( \frac{\partial \psi}{\partial z} +\tan\delta  \frac{\partial \psi}{\partial y} \right) + \frac{D_y}{2k_0 n_0} \frac{\partial^2 \psi}{\partial y^2} + \frac{k_0}{2n_0} \left( n_e^2 - n_0^2 \right) \psi \nonumber \\
			+ \frac{k_0 c}{2} \left(P_\mathrm{NL,y}-\tan\delta\ P_\mathrm{NL,z} \right) - i\frac{c D_y}{2 n_0} \frac{\partial P_\mathrm{NL,z}}{\partial y}	= 0,  \label{eq:NLSE_paraxial} 
\end{align}
where $c$ is the speed of light in vacuum, $\psi=H_x e^{-i k_0 n_0 z}$ is the slowly varying envelope and $n_0=n_e(\xi)$.\\
In the linear regime ($\bm{P}_\mathrm{NL}=0$), Eq.~\eqref{eq:NLSE_paraxial} is a Schr\"odinger-like equation for a massive particle of charge \textsl{e} in an electromagnetic field:
\begin{equation}
i\hbar\frac{\partial \psi}{\partial t} =\frac{\left[i\hbar\nabla-(e/c)\bm{A}\right]^2}{2m} \psi+ U \psi.
\label{eq:schrodinger} 
\end{equation}
The Hamiltonian corresponding to Eq.~\eqref{eq:schrodinger} is $\hat{H}=\frac{\left[\bm{\hat{p}}- (e/c)\bm{\hat{A}}\right]^2}{2m} + \hat{U}$, where $\bm{A}$ and $U$ are the vector and scalar electromagnetic potentials, respectively. To transform Eq.~\eqref{eq:schrodinger} into \eqref{eq:NLSE_paraxial} we need to carry out the transformations $t\rightarrow z$, $\bm{p}\rightarrow k_y \hat{y}$, $\hbar\rightarrow 1$, $\bm{A}\rightarrow -\frac{k_0 n_0 \tan\delta}{D_y} \hat{y}$, $m\rightarrow \frac{k_0 n_0}{D_y}=\frac{k_0\epsilon_{zz}}{n_0}$ and $U\rightarrow -\frac{k_0\left(n_e^2-n_0^2\right)}{2n_0}- \frac{k_0 n_0 \tan^2\delta}{D_y}$ \footnote{We stress once again that the quantities $n_0=n_e(\xi)$, $\delta(\xi)$ and $D_y(\xi)$ in Eq.~\eqref{eq:NLSE_paraxial} are computed for $k_y=0$ (i.e., $\theta=\xi$) and their spatial variations are neglected.}. Thus, light propagation in hyperbolic materials resembles the motion of a particle with negative mass. \\
Such an effective mass in HMs of type I is plotted in Fig.~\ref{fig:effective_mass}(a): it is always negative (like $D_y$) and increases monotonically with $\theta$, vanishing at the edge of the existence region despite $D_y\rightarrow\infty$. Hence, the beam diffraction is expected to increase with $\theta$, as we verified by computing the solutions of Eq.~\eqref{eq:NLSE_paraxial} in the linear regime and in the homogeneous case using a plane-wave expansion \cite{Paul:2009}. As illustrated in Fig.~\ref{fig:effective_mass}(b), the beam spreading markedly increases as $\theta$ gets larger, with walk-off corresponding to a plane wave as in Fig.~\ref{fig:comparison_diffraction}, even though the beam is a few wavelengths in size. 

\section{Negative refraction}
\subsection{Particle-like model}
The analogy drafted above between light propagation in HMs and the motion of a charged particle of negative mass provides a simple explanation for negative refraction at the interface between an isotropic material and an HM \cite{Smith:2004,Yao:2008}. We consider Eq.~\eqref{eq:NLSE_paraxial} in a framework with axis $z$ normal to the interface [Fig.~\ref{fig:sketch}(c)]. Thus, the paraxial approximation will  be rigorously valid only at normal incidence, as the quantities $n_e$, $D_y$ and $\delta$ were computed for phase fronts normal to $\hat{z}$, that is, $\theta=\xi$. \\
 The effective (transverse) velocity is $v=dy/dz$, corresponding to the tangent of the angle $\alpha$ formed by the ray with $\hat{z}$. For a system invariant across $y$, the canonical momentum $p$ (the transverse component of the wave vector) is conserved when light waves cross the interface, providing for the velocity (i.e., the direction of the energy flux vector)
\begin{equation}   \label{eq:momentum_conservation}
  \left. \frac{d \left\langle y \right\rangle}{dz} \right|_{hyperbolic} = D_y \frac{n_{iso}}{n_0} \left. \frac{d \left\langle y \right\rangle}{dz} \right|_{isotropic} + \tan\delta,
\end{equation}
with $n_{iso}$ the refractive index of the isotropic material and $\left\langle y \right\rangle=\int{y \left| \psi\right|^2dy}/ \int{\left| \psi\right|^2dy}$. The first term on the RHS of Eq.~\eqref{eq:momentum_conservation} accounts for the role of dispersion in determining the longitudinal component of the wave vector, as $k_y$ is dictated by the boundary condition \cite{Smith:2004}. It results in a refracted beam which is flipped with respect to the axis $z$, therefore, it undergoes negative refraction. The second term on the RHS of Eq.~\eqref{eq:momentum_conservation} is the walk-off contribution and quantifies the angular deviation of the reference system $x^\prime y ^\prime z^\prime$ from $xyz$ in the plane $yz$. Naming $\alpha_\mathrm{in}=\arctan{\left(\left. \frac{d \left\langle y \right\rangle}{dz} \right|_{isotropic} \right)}$ the incidence angle of the impinging beam, with $\alpha_\mathrm{in}=\beta_\mathrm{in}$ owing to isotropy of the first medium, negative refraction always occurs when $\beta_\mathrm{in} \delta<0$, but it requires $\left|\tan\delta \right| < - \frac{n_{iso} D_y}{n_0} \left| \tan{\beta_\mathrm{in}} \right|$ when  $\beta_\mathrm{in} \delta>0$. 

\subsection{Comparison between particle model and exact solutions}
\begin{figure}
\includegraphics[width=0.5\textwidth]{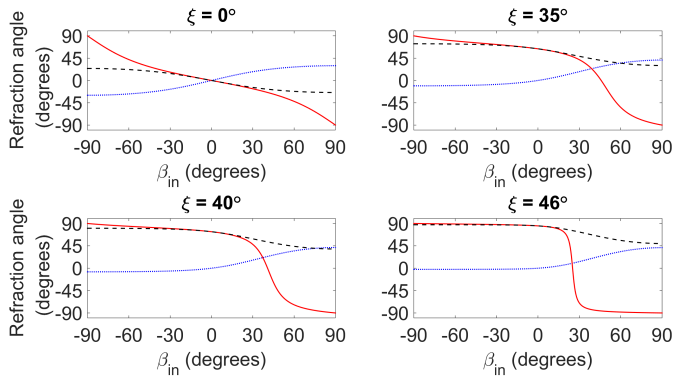}
\caption{\label{fig:negative_refraction} Negative refraction in a Type I HM for four different orientations $\xi$ of the optic axis (as marked). Wave vector refraction angle $\beta$ (dotted blue lines) and  angle  $\alpha$ (black dashed lines) versus incidence angle $\beta_\mathrm{in}$ as computed from Eq.~\eqref{eq:beta} and Eq.~\eqref{eq:alpha}, respectively. Red solid lines graph Poynting vector refraction resulting from the particle-like model Eq.~\eqref{eq:momentum_conservation}. Here $\epsilon_\bot=2.2614$ and $\epsilon_\|=-2.8744$.}
\end{figure}

It is worthwhile to validate Eq.~\eqref{eq:momentum_conservation}, in the paraxial approximation, against the exact solutions. In the plane wave limit, from Eqs.~(\ref{eq:maxwell_y_2}-\ref{eq:maxwell_x_2}) we can derive the dispersion relation
\begin{equation}
  \left(k_z +\tan\delta\ k_y \right)^2 + D_{y} k_y^2 = k_0^2 n_{0}^2 . 
	\label{eq:dispersion}
\end{equation}
Clearly, in the plane wave limit the two coefficients $\delta$ and $D_y$ account for the spatial dispersion stemming from anisotropy.\\
To find the angle of refraction we need to know the relation $k_z=k_z(k_y)$, where $k_y=k_0 n_{iso} \sin \beta_\mathrm{in}$ is the transverse component of the incident wave vector. Here we deal with forward waves only, with $-\pi/2<\beta<\pi/2$ \footnote{Reciprocity ensures a symmetric response with respect to forward and backward waves.}. From Eq.~\eqref{eq:dispersion}, the angle $\beta$ of the wave vector after refraction is
\begin{equation}
 \beta(\beta_\mathrm{in},\xi)=\arctan{\left( \frac{1}{-\tan\delta \pm \sqrt{\frac{n_0^2}{n_{iso}^2\sin^2\beta_\mathrm{in}}-D_{y}}} \right)}. \label{eq:beta}
\end{equation}
According to Eq.~\eqref{eq:beta}, if $n_0$ is real (the latter implies that beams impinging normally to the interface do not excite evanescent waves), a real angle $\beta$ exists for any incidence angle $\beta_\mathrm{in}$, as $D_y<0$ [see Eq.~\eqref{eq:diffraction_coefficient}]. The angle $\beta$ is plotted versus $\beta_\mathrm{in}$ in Fig.~\ref{fig:negative_refraction} (blue dotted line) for various $\xi$. The sign in front of the square root in Eq.~\eqref{eq:beta} must be chosen in order to get positive refraction for the wave vector via the conservation of $k_y$ at the interface, i.e., $\beta \beta_\mathrm{in}>0$ \cite{Smith:2004}. For arbitrary orientations $\xi$ of the optic axis, at normal incidence ($\beta_\mathrm{in}=0$) it is $\beta=0$, as required by momentum conservation. For $\xi=0$ light propagation obeys mirror symmetry with respect to $z$; however, for $\xi\neq 0$ refraction becomes non-specular with respect to  left/right inversion, with $\beta$ getting larger when the incident beam is tilted on the same side of the optic axis, i.e., $\beta_\mathrm{in}\xi>0$. \\  
The direction of the refracted Poynting vector is obtained by adding the walk-off angle to $\beta$:
\begin{equation}
  \alpha(\beta_\mathrm{in},\xi)=\beta + \arctan{ \left\{ \frac{\epsilon_a \sin \left[ 2(\xi-\beta)\right]}{\epsilon_a + 2\epsilon_\bot + \epsilon_a \cos\left[2\left(\xi-\beta\right)\right]} \right\}}, \label{eq:alpha}
\end{equation} 
as plotted by dashed black lines in Fig.~\ref{fig:negative_refraction}.

When $\xi=0$, the walk-off has opposite sign with respect to $\beta$ (see Fig.~\ref{fig:comparison_diffraction}) and is about twice larger in modulus: hence, the energy flow is negatively refracted for any incidence angle. When $\xi\neq 0§$, refraction is always negative  when $\xi\beta_\mathrm{in}<0$, whereas for $\xi\beta_\mathrm{in}>0$ negative refraction occurs only for small $\xi$ (not shown), and for incidence angles $\beta_\mathrm{in}$ below a threshold  depending on $n_{iso}$ and the eigenvalues of $\bm{\epsilon}$. For $\xi$ exceeding the existence cone defined by Eq.~\eqref{eq:typeI}, the RHS of Eq.~\eqref{eq:beta} is no longer real, and homogeneous (i.e., non-evanescent) waves exist only in some narrow intervals of $\beta_\mathrm{in}$. Here we are not interested in such solutions.\\
Having described ``exactly'' refraction at the interface between an isotropic medium and a type I HM, we can now address the accuracy of the particle-like (paraxial) model. Figure~\ref{fig:negative_refraction} graphs the refraction of the Poynting vector (red solid lines) given by Eq. \eqref{eq:momentum_conservation} in the framework of the particle model. The results from particle-like and exact models are in very good agreement for $\left|\beta_\mathrm{in}\right|<30^\circ$, consistently with the paraxial approximation. For $\xi\neq 0^\circ$, owing  to anisotropy the discrepancy between them depends on the sign of the incidence angle, with bigger differences for larger incidence angles (absolute values). Based on Eq.~\eqref{eq:momentum_conservation}, negative refraction always occurs for $\left|\beta_\mathrm{in}\right|\rightarrow 90^\circ$, despite the orientation of the optic axis (i.e., the value of $\xi$). In fact, $\tan\delta$ is the dominant term on the RHS of Eq.~\eqref{eq:momentum_conservation} for large  $\left|\beta_\mathrm{in}\right|$; hence, negative refraction is expected for both positive and negative incidence angles, with the refracted beam eventually propagating at grazing angles according to Fig.~\ref{fig:comparison_diffraction}. The transition between positive and negative $\alpha$ becomes very steep as the HM approaches the limit \eqref{eq:typeI}, corresponding to a singularity in the coefficient $D_y(\theta)$.

\subsection{Numerical simulations}

\begin{figure}
\includegraphics[width=0.5\textwidth]{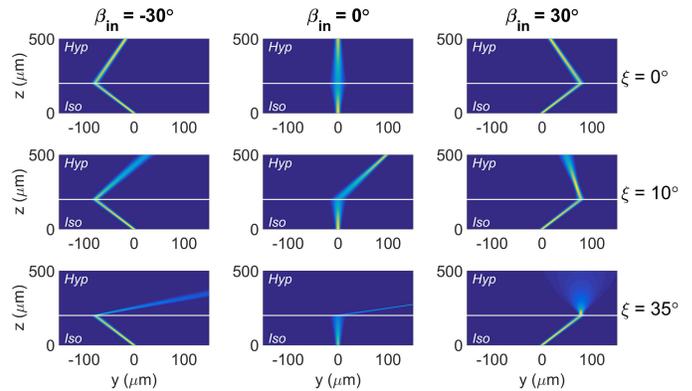}
\caption{\label{fig:BPM_negative_refraction} Negative refraction in a Type I HM with $n_{iso}=1$, $\epsilon_\bot=2.2614$, $\epsilon_\|=-2.8744$ and $\lambda=1064$nm. Evolution in the plane $yz$ for an input Gaussian beam of waist 5$\mu$m for various incidence angles $\beta_\mathrm{in}$ (columns) and optic axis orientations $\xi$ (rows), as labelled. The white solid lines represent the boundary between the isotropic medium (lower region) and the HM (upper part).}
\end{figure}

\begin{figure}
\includegraphics[width=0.5\textwidth]{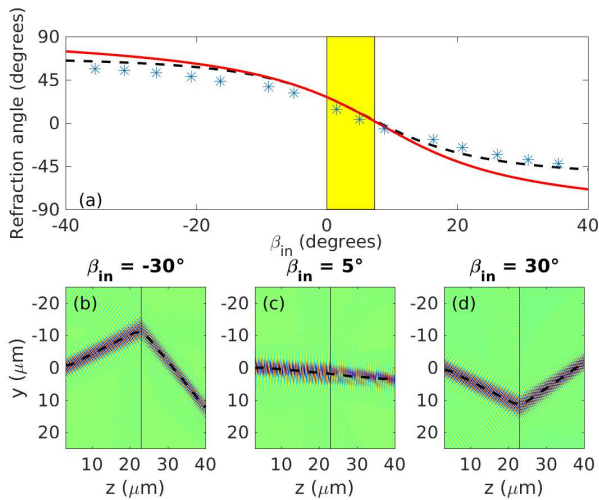}
\caption{\label{fig:FDTD_negative_refraction} Light refraction at the interface isotropic-hyperbolic  when the optic axis is at  $\xi= 5^\circ$. (a) Refraction angle $\alpha$ versus incidence angle $\beta_\mathrm{in}$ according to FDTD  (points), exact values for plane waves [dashed black line, Eq.~\eqref{eq:alpha}] and paraxial approximation [red solid line, Eq.~\eqref{eq:momentum_conservation}]. The yellow shaded rectangle marks the region with positive refraction. (b) Snapshots of the $y$-polarized electric field distribution in the propagation plane $yz$ for (b) $\beta_\mathrm{in}=-30^\circ$, (c) $\beta_\mathrm{in}=5^\circ$ and (d) $\beta_\mathrm{in}=30^\circ$, respectively. In (b-d) the dashed lines plot the beam trajectory, computed as the intensity peak in each section normal to $z$. In the FDTD simulations we considered $\epsilon_\bot = 1.755$, $\epsilon_\| = -0.3805 - 0.0299i$ [the imaginary part is neglected both in Eqs.~\eqref{eq:alpha} and \eqref{eq:momentum_conservation}]. Here the wavelength is $0.83\mu$m. }
\end{figure}

We checked the validity of the plane-wave results illustrated in Fig.~\ref{fig:negative_refraction} in the case of finite beams by using a beam propagation (BPM) code and a finite-difference time-domain (FDTD) open source simulator, MEEP \cite{Oskooi:2010}. The BPM provides accurate results only when the paraxial approximation is applicable, the FDTD code does not have such limitation. The BPM results are plotted in Fig.~\ref{fig:BPM_negative_refraction}, where we carried out our simulations below a maximum incidence  $\left|\beta_\mathrm{in}\right| = 30^\circ$, compatible with the paraxial approximation. At normal incidence $\beta_\mathrm{in}=0^\circ$, the beam at the interface undergoes a deflection corresponding to the walk-off angle graphed in Fig.~\ref{fig:comparison_diffraction} with $\theta=\xi$. Beam refocusing is observed inside the hyperbolic medium owing to negative diffraction \cite{Smith:2004}. When the input beam is tilted to the other side of the optic axis (i.e., $\xi\beta_\mathrm{in}<0$), negative refraction always occurs, the larger $\xi $ the larger the angle $\alpha$ of the Poynting vector. Conversely, when input beam and optic axis are directed on the same side, the magnitude of negative refraction decreases with $\xi$ and its range  of occurrence $\beta_\mathrm{in}$ reduces, consistently with the non-paraxial model [see Eq.~\eqref{eq:alpha} and Fig.~\ref{fig:negative_refraction}]. The paraxial model \eqref{eq:momentum_conservation}, conversely, unphysically predicts negative refraction for large positive incidence angles (assuming $\xi\beta_\mathrm{in}>0$, with the same behavior in the opposite case after a specular reflection about $\hat{z}$) (see Fig.~\ref{fig:negative_refraction}). When the input wave vector approaches the boundaries of the existence cone Eq.~\eqref{eq:typeI}, the validity range of the paraxial model narrows towards the right edge (for positive $\xi$): for example, for $\xi=35^\circ$ and $\beta_\mathrm{in}=30^\circ$ (last panel in Fig.~\ref{fig:BPM_negative_refraction}), beam refraction appears borderline between negative and positive ($\alpha\approx 0$), but refraction should be positive based on the exact solution (see Fig.~\ref{fig:negative_refraction}). \\
Figure~\ref{fig:FDTD_negative_refraction} illustrates FDTD simulations of light behavior at air-HM interface for various incidence angles $\beta_\mathrm{in}$ and $\xi=5^\circ$. The angle of refraction from FDTD closely follows the predictions of \eqref{eq:alpha}, the latter  rigorously valid  for plane waves. Noteworthy, the FDTD match more closely  \eqref{eq:alpha} than the paraxial model \eqref{eq:momentum_conservation}, with small discrepancies between FDTD and plane wave model mainly due to the presence of losses [neglected in Eq.~\eqref{eq:alpha}]. Losses, even if small, affect light propagation in a non-negligible way \cite{Paul:2009,Yu:2016}. For instance, at normal incidence ($\beta_\mathrm{in}=0^\circ$ and thus $\alpha=\delta$), a change of about $10^\circ$ on the trajectory slope is visible [Fig.~\ref{fig:FDTD_negative_refraction}(a)], such variation being exclusively due to a difference in the walk-off angle. Generally, losses decrease the absolute value of $\alpha$ (see Fig.~\ref{fig:FDTD_negative_refraction}). In agreement with the theoretical predictions, for small positive rotations of the optic axis $\xi$ light undergoes negative refraction at the interface, except for a narrow interval limited by $\beta_\mathrm{in}=0^\circ$ and an upper extremum depending on $\xi$ (yellow shaded region in Fig.~\ref{fig:FDTD_negative_refraction}).

\section{Diffraction compensation}
\begin{figure}
\includegraphics[width=0.5\textwidth]{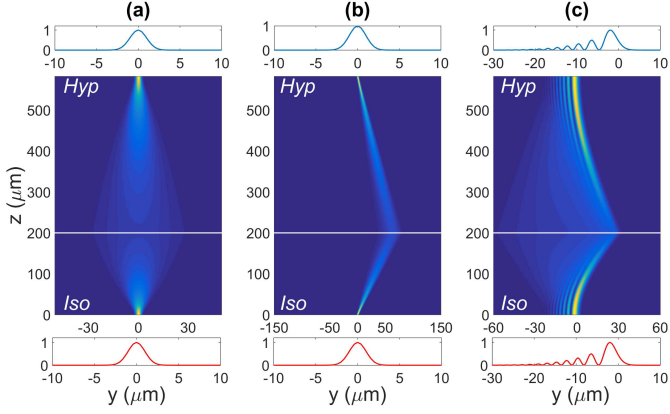}
\caption{\label{fig:compensation_diffraction} Diffraction compensation in a two-layer stack with an isotropic medium (bottom, $ L_{iso}=200\mu$m, $n_{iso}=1$) and a Type I HM (top, $\xi=0$, $\epsilon_\bot=2.2614$, $\epsilon_\|=-2.8744$) for a beam of wavelength  $\lambda=1.064 \mu$m. The white line indicates the interface between the two media. (a) Evolution of an input Gaussian beam with waist 2$\mu$m and planar phase-front; (b) same as in (a) but with phase-front tilted by 20$^\circ$; (c) evolution of an Airy beam modulated by a Gaussian beam, input profile $\psi_0=\text{Ai}(y/w_A)\exp{\left(-2y^2/w_G^2\right)}$ with $w_A=2\mu$m and $w_G=20\mu$m. Top and bottom graphs show input and output profiles, respectively. 
}
\end{figure}

The possibility of beam anomalous diffraction  entails the realization of spatial equivalents of dispersion-compensators with temporal pulses in optical fibers, where opposite signs of chromatic dispersion are alternated \cite{Jopson:1995}. Beam diffraction in an isotropic slab of length $L_{iso}$ and refractive index $n_{iso}$ can be canceled out by subsequent propagation in an HM of extent 
\begin{equation}   \label{eq:Lhyp}
  L_{hyp}= \frac{n_0}{n_{iso} \left|D_y \right|} L_{iso}.
\end{equation}
If $\psi_{0}$ is the input beam profile in $z=0$ and Eq.~\eqref{eq:Lhyp} is satisfied, at the output of the second slab (HM), after a propagation length $L_{iso}+L_{hyp}$, we expect to find a replica of $\psi_0$. Figure~\ref{fig:compensation_diffraction} demonstrates this concept via BPM simulations, launching three distinct beam profiles $\psi_0$ in such a two-layer structure. It can be seen that the output profiles coincide with the input, even though our model relies on the paraxial approximation, without resorting to superlens effects based on the recovery of evanescent waves \cite{Pendry:2000,Jacob:2006}. For Type I HM and $\theta=0$, the side-shift due to walk-off  vanishes (Fig.~\ref{fig:comparison_diffraction}) and the replica retrieves its transverse position at the input. Conversely, in the presence of walk-off, the output field is laterally shifted by $L_{hyp}\tan\delta$. It needs to be underlined that anomalous diffraction can also occur in photonic crystals \cite{Kosaka:1998} and waveguide arrays \cite{Eisenberg:2000}, but in both these systems the field is an envelope of Bloch waves, thus in general a faithful reconstruction of the input profile is inhibited. On the contrary, in hyperbolic media such limitation is not present, as long as the effective medium theory remains valid \cite{Zhang:2015}. In the case of hyperbolic metamaterials, the subwavelength unit cell fixes the minimum resolution achievable. Before achieving this limit, spatial nonlocal effects have to be accounted for \cite{Silveirinha:2009,Rytov:1956,Orlov:2011}. Physically, the field reconstruction in the present structure can be interpreted as a time inversion occurring in the hyperbolic slab while conserving the sign of the effective mass, i.e., a shift of the minus sign from the RHS to the LHS of  Eq.~\eqref{eq:schrodinger} \cite{Smolyaninov:2015}.  Since in real media the propagation losses have to be accounted for, as they can strongly affect diffraction \cite{Paul:2009}, in the following subsection we will address this issue using FDTD simulations.

\subsection{FDTD analysis}
\begin{figure}
\includegraphics[width=0.5\textwidth]{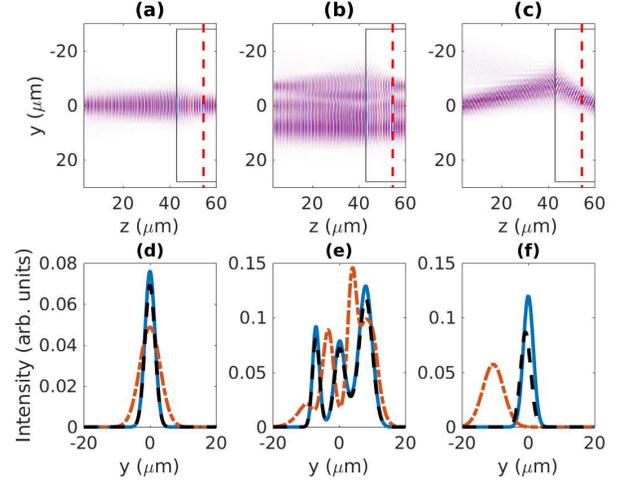}
\caption{\label{fig:compensation_diffraction_FDTD} (Color online) FDTD results demonstrating  diffraction compensation for $\xi=0^\circ$. Snapshot of the electric field after reaching the stationary regime, showing diffraction compensation (top) for (a) a  bell-shaped input, (b) a three-hump input and (c) a bell-shaped input impinging on the interface at an angle of $10^\circ$. Bottom: corresponding transverse profiles of the input intensity (blue solid lines), of the output without (orange dash-dotted lines) and with (dashed black lines) the HM slab; the output section is in $z=60 \mu$m (at the end of the grid). The dashed red lines in (a-c) represent the analytically calculated length $L_{hyp}$. Here $\lambda=0.83\mu$m, $L_{iso}=40 \mu$m, $n_{iso}=1$, $L_{hyp}=11.5 \mu$m (dashed vertical lines), $\epsilon_\bot = 1.755$, $\epsilon_\| = -0.3805 - 0.0299i$.}
\end{figure}

When considering an actual hyperbolic material, losses can strongly affect light propagation \cite{Paul:2009,Argyropoulos:2013}. For instance, when modeling the medium polarizability with the Lorentz oscillator, a negative permittivity is expected in the proximity of an absorption line. Stated otherwise, the Kramers-Kronig relations do not allow to set independently the real and imaginary parts of the dielectric permittivity \cite{Silveirinha:2009}. We used the  MEEP FDTD program \cite{Oskooi:2010} and considered the case $\xi=0^\circ$, corresponding to vanishing walk-off, in order to underline the role of  negative diffraction. Figure~\ref{fig:compensation_diffraction_FDTD} shows light propagation for three different input profiles and moderate propagation losses. The single-hump beam undergoes refocusing when entering  the hyperbolic material, in agreement with our analytical predictions and BPM simulations [see Fig.~\ref{fig:compensation_diffraction}(a)]. When the input is tilted with respect to the interface normal $\hat{z}$, the beam undergoes negative refraction and eventually retrieves its original transverse position, nearly recovering its input profile through anomalous diffraction [see Fig.~\ref{fig:compensation_diffraction_FDTD}(c)]. When a three-humps beam is launched, the propagation in the HM allows forming a mirror image of the input [see Fig.~\ref{fig:compensation_diffraction_FDTD}(b)]. These FDTD numerical experiments demonstrate that the main features of the phenomenon survive moderate losses. Nonetheless, besides the inevitable reduction in power, the mirror plane (image) gets shifted more than predicted inside the lossy HM slab. If we define $L_{hyp}^{real}$ the distance of the image plane from the interface air-HM in the presence of losses, we get a relative difference of $(L_{hyp}^{real}-L_{hyp})/L_{hyp}^{real}\approx 33\%$ in the case plotted in Fig.~\ref{fig:compensation_diffraction_FDTD}.  

\section{Propagation in a graded-index}
 
\begin{figure}
\includegraphics[width=0.5\textwidth]{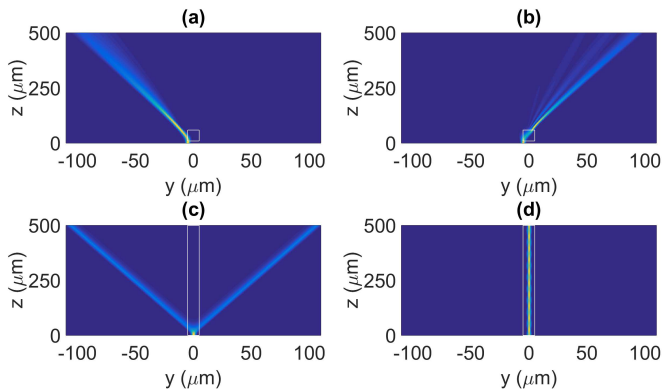}
\caption{\label{fig:BPM_GRIN} Interaction of a $\lambda=1.064 \mu$m $2\mu$m-wide Gaussian beam and flat phase fronts with a GRIN distribution in a Type I HM characterized by $\xi=0$, $\epsilon_\bot=2.2614$, $\epsilon_\|=-2.8744$. (a-b) Beam evolution in the plane $yz$ for input in $y=-5\mu$m  and normal incidence, in the presence of a Gaussian $y$-graded waveguide segment located along $z$ between $z=10 \mu$m and $z=50 \mu$m, of width $w_\mathrm{GRIN}=5\mu$m and (a) $\left(\Delta n\right)_0 = 0.1$  and (b) $\left(\Delta n\right)_0 = -0.1$. (c-d) As in (a-b), but with the input beam in $y=0\mu$m and a $z$-extended waveguide. The white rectangles mark the location of the GRIN regions.}
\end{figure}

An effective negative mass can yield exotic interactions of light with graded distributions of refractive index (GRIN). As stated by the Fermat's principle, in media with elliptic dispersion light is ``attracted'' towards regions with a higher refractive index. This property can be reformulated in the framework of the Schr\"{o}dinger equation, where the optical analog of the Ehrenfest's theorem states that beams are subject to a transverse force proportional to the sign-inverted transverse gradient of the refractive index $n$, $-\nabla_\mathrm{t} n$. Since the effective mass of light beams is negative in hyperbolic media, despite the equivalent force is always anti-parallel to the index gradient, the beam gets ``pulled'' towards lower refractive index regions. Examples of beam interactions with $y$-dependent 
Gaussian GRIN distributions $n_e(y)-n_0=\left(\Delta n\right)_0 e^{-y^2/w_\mathrm{GRIN}^2}$ are presented in Fig.~\ref{fig:BPM_GRIN}(a-b). The effective negative mass  causes the beam deviation to flip as compared with standard materials encompassing elliptic dispersion. Similarly, GRIN waveguides in HMs require a lower refractive index in the core than in the cladding, in contrast  to standard waveguides [see Fig.~\ref{fig:BPM_GRIN}(c-d)].  

\section{Nonlinear case}
Here we address the role of a third-order nonlinear response such as an intensity-dependent refractive index. For the sake of simplicity, we refer to a standard Kerr material by setting $n_e-n_0=n_{2H}\left|H_x\right|^2$, i.e., an index change proportional to the local electromagnetic intensity. In terms of the standard Kerr coefficient $n_{2E}$ referred to the square of the electric field, it is $n_{2E}Z^2=n_{2H}$, with $Z$ the medium impedance. The inverted profile of a confining waveguide in HMs suggests that, at variance with elliptic dispersion, bright or dark solitons are supported by negative ($n_{2H}<0$) or positive ($n_{2H}>0$) nonlinearities, respectively, in analogy to temporal solitons in fibers \cite{Hasegawa:1973,Hasegawa:1973_1}.\\
\begin{figure}
\includegraphics[width=0.5\textwidth]{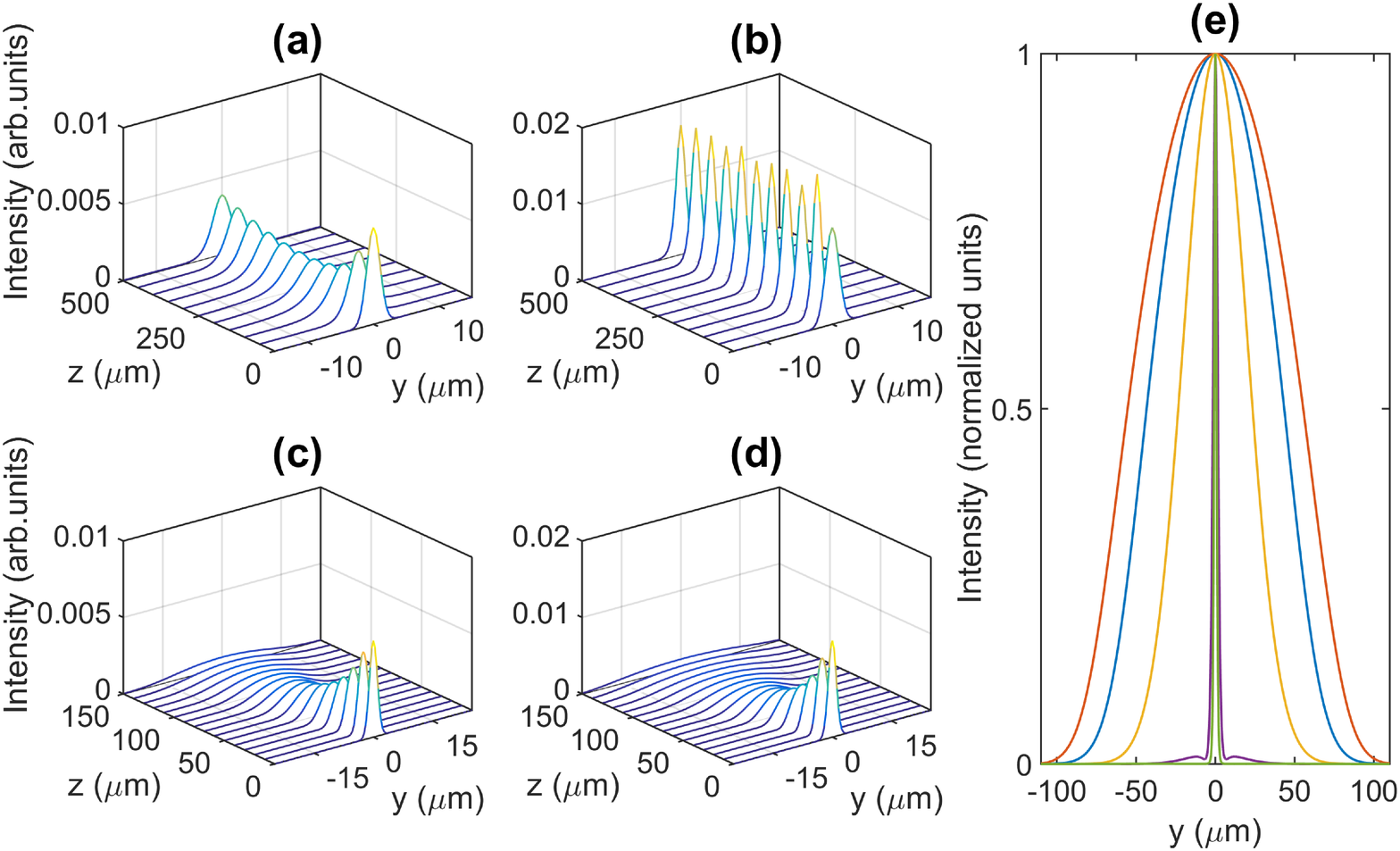}
\caption{\label{fig:nonlinear_case} Nonlinear evolution of a Gaussian beam of waist $2\mu$m, power $P$, $\lambda=1.064 \mu$m and planar phase front at the input $z=0$ in a Type I HM with $\xi=0$, $\epsilon_\bot=2.2614$, $\epsilon_\|=-2.8744$. Beam cross-sections versus $z$ for (a) $n_{2E}P=-5\times10^{-12}$m$^2$V$^{-2}$W, (b) $n_{2E}P=-1\times10^{-11}$m$^2$V$^{-2}$W, (c) $n_{2E}P=5\times10^{-12}$m$^2$V$^{-2}$W and (d) $n_{2E}P=1\times10^{-11}$m$^2$V$^{-2}$W. (e) Comparison of the normalized intensity distributions in $z=500\mu$m  versus $y$: from the widest to the narrowest, the excitations correspond to the cases (d) (red line), (c) (blue), linear  (yellow), (a) (magenta) and (b) (green), respectively.}
\end{figure}
Figure~\ref{fig:nonlinear_case} shows the BPM-computed evolution of a  Gaussian beam input in either a focusing ($n_{2H}<0$) [Fig.~\ref{fig:nonlinear_case}(a-b)] or a defocusing ($n_{2H}>0$) [Fig.~\ref{fig:nonlinear_case}(c-d)] HM. As expected, when $n_{2H}<0$ the Gaussian beam evolves into a fundamental single-humped soliton featuring a hyperbolic secant profile, emitting radiation while adjusting to the stationary state, in agreement with inverse scattering theory \cite{Kivshar:2003}. When $n_{2H}>0$ the beam retains its bell shape, but spreads more than in the linear case $n_{2H}=0$ [see Fig.~\ref{fig:nonlinear_case}(e)]. While this counter-intuitive behavior was partially discussed in Refs.~\cite{Kou:2011,Silverinha:2013} (plasmonic waveguide arrays with nanowires) and \cite{Smolyaninov:2013} (analogy with gravitational forces between photons), our model provides a markedly simpler and physically intuitive explanation in terms of anomalous diffraction, retaining its validity in a variety of systems and materials, including, e.g., natural HMs \cite{Sun:2014,Esslinger:2014}. 

\section{Summary and outlook}

We modeled light propagation in materials with hyperbolic dispersion as the motion of a quantum particle possessing a negative mass \cite{DiMei:2016}. A negative effective mass corresponds to anomalous diffraction and provides a straightforward explanation of negative refraction at the interface between hyperbolic and isotropic media. We compared our results in the paraxial approximation with exact solutions in order to address their range of applicability. We found an explicit closed-form for the angle of refraction  [Eq.~\eqref{eq:beta}],  generally applicable to refraction from an isotropic material to a uniaxial in the case of  co-planar optic axis and input wave vector. Through time-inversion of light propagation in a homogeneous HM, our model allows designing novel structures for the perfect reconstruction of arbitrary paraxial input fields. Compared with classical configurations based on lenses (e.g. the 4f correlator), HM-based reconstructions are invariant with respect to transverse shifts of the beam, representing an ideal design for short distance optical communications based upon spatial multiplexing \cite{Zhao:2015}. Our results demonstrate that complex waveforms can be faithfully retrieved, even in the presence of moderate losses - unavoidable due to Kramers-Kronig relations \cite{Argyropoulos:2013} - and with spatial resolution determined by the material nonlocality \cite{Poddubny:2013,Rytov:1956,Orlov:2011}. The  negative effective mass of light implies attraction towards (repulsion from) regions with a lower (higher) refractive index, opposite to the standard behavior when dispersion is elliptic; hence, in the Kerr (cubic) regime, spatial bright (dark) solitons are supported by a negative (positive) intensity-dependent refractive index. 
Finally, through reciprocity between magnetic and electric properties, our results are also valid in magnetic hyperbolic metamaterials \cite{Kruk:2016} as well as in the presence of more complex bi-anisotropic responses \cite{Smith:2003}.  


\section*{Acknowledgments} A.A. and G.A. thank the Academy of Finland through the Finland Distinguished Professor grant no. 282858. C.P.J. gratefully acknowledges Funda\c{c}\~{a}o para a Ci\^{e}ncia e a Tecnologia, POPH-QREN and FSE (FCT, Portugal) for the fellowship SFRH/BPD/77524/2011; she also thanks the Optics Lab in Tampere for hospitality.

%

\end{document}